\documentclass[prb,superscriptaddress,floatfix,a4paper,aps,twocolumn,nofootinbib]{revtex4-1}
\usepackage[utf8]{inputenc}
\usepackage{natbib}
\usepackage{amsmath}
\usepackage{amsbsy}
\usepackage{amssymb}
\usepackage{scrextend}
\usepackage{graphicx}
\usepackage{color}
\usepackage[usenames, dvipsnames]{xcolor}
\usepackage[colorlinks=true, allcolors = JungleGreen]{hyperref}

\renewcommand{\vec}{\boldsymbol}
\newcommand{\sss}{\scriptscriptstyle}

\begin{document}
\title{Thermal creep induced by cooling in a superconducting vortex lattice}
\author{Roland Willa}
\affiliation{Institute for Theoretical Physics, ETH Zurich, 8093 Zurich, Switzerland}
\affiliation{Materials Science Division, Argonne National Laboratory, Argonne, Illinois 60439, USA}
\affiliation{Institute of Condensed Matter Theory, Karlsruhe Institute of Technology, 76131 Karlsruhe, Germany}
\author{Jose Augusto Galvis}
\affiliation{Departamento de Ciencias Naturales, Facultad de Ingenier\'ia y Ciencias B\'asicas, Universidad Central, Bogot\'a, Colombia}
\author{Jose Benito-Llorens}
\affiliation{Laboratorio de Bajas Temperaturas y Altos Campos Magn\'eticos, Unidad Asociada UAM/CSIC, Departamento de F\'isica de la Materia Condensada, Instituto de Ciencia de Materiales Nicol\'as Cabrera, Condensed Matter Physics Center (IFIMAC), Universidad Aut\'onoma de Madrid, E-28049 Madrid, Spain}
\author{Edwin Herrera}
\affiliation{Departamento de Ciencias Naturales, Facultad de Ingenier\'ia y Ciencias B\'asicas, Universidad Central, Bogot\'a, Colombia}
\affiliation{Laboratorio de Bajas Temperaturas y Altos Campos Magn\'eticos, Unidad Asociada UAM/CSIC, Departamento de F\'isica de la Materia Condensada, Instituto de Ciencia de Materiales Nicol\'as Cabrera, Condensed Matter Physics Center (IFIMAC), Universidad Aut\'onoma de Madrid, E-28049 Madrid, Spain}
\author{Isabel Guillamon}
\affiliation{Laboratorio de Bajas Temperaturas y Altos Campos Magn\'eticos, Unidad Asociada UAM/CSIC, Departamento de F\'isica de la Materia Condensada, Instituto de Ciencia de Materiales Nicol\'as Cabrera, Condensed Matter Physics Center (IFIMAC), Universidad Aut\'onoma de Madrid, E-28049 Madrid, Spain}
\author{Hermann Suderow}
\affiliation{Laboratorio de Bajas Temperaturas y Altos Campos Magn\'eticos, Unidad Asociada UAM/CSIC, Departamento de F\'isica de la Materia Condensada, Instituto de Ciencia de Materiales Nicol\'as Cabrera, Condensed Matter Physics Center (IFIMAC), Universidad Aut\'onoma de Madrid, E-28049 Madrid, Spain}
\date{\today}

\begin{abstract}

A perturbed system relaxes towards an equilibrium given by a minimum in the potential energy landscape. This often occurs by thermally activated jumps over metastable states. The corresponding dynamics is named creep and follows Arrhenius' law. Here we consider the situation where the equilibrium position depends on temperature. We show that this effect occurs in the vortex lattice of the anisotropic superconductor $2\mathrm{H}$-$\mathrm{Nb}\mathrm{Se}_{2}$ when the magnetic field is tilted away from the principal axes, and that it leads to the peculiar appearance of creep when cooling the sample. Temperature determines the system's ground state and at the same time brings the system back to equilibrium, playing a dual and antagonistic role. We expect that cooling induced creep occurs in correlated systems with many degrees of freedom allowing to tune the equilibrium state via heat treatment.
\end{abstract}

\maketitle

\begin{figure}[t]
\centering
\includegraphics[width=0.48 \textwidth]{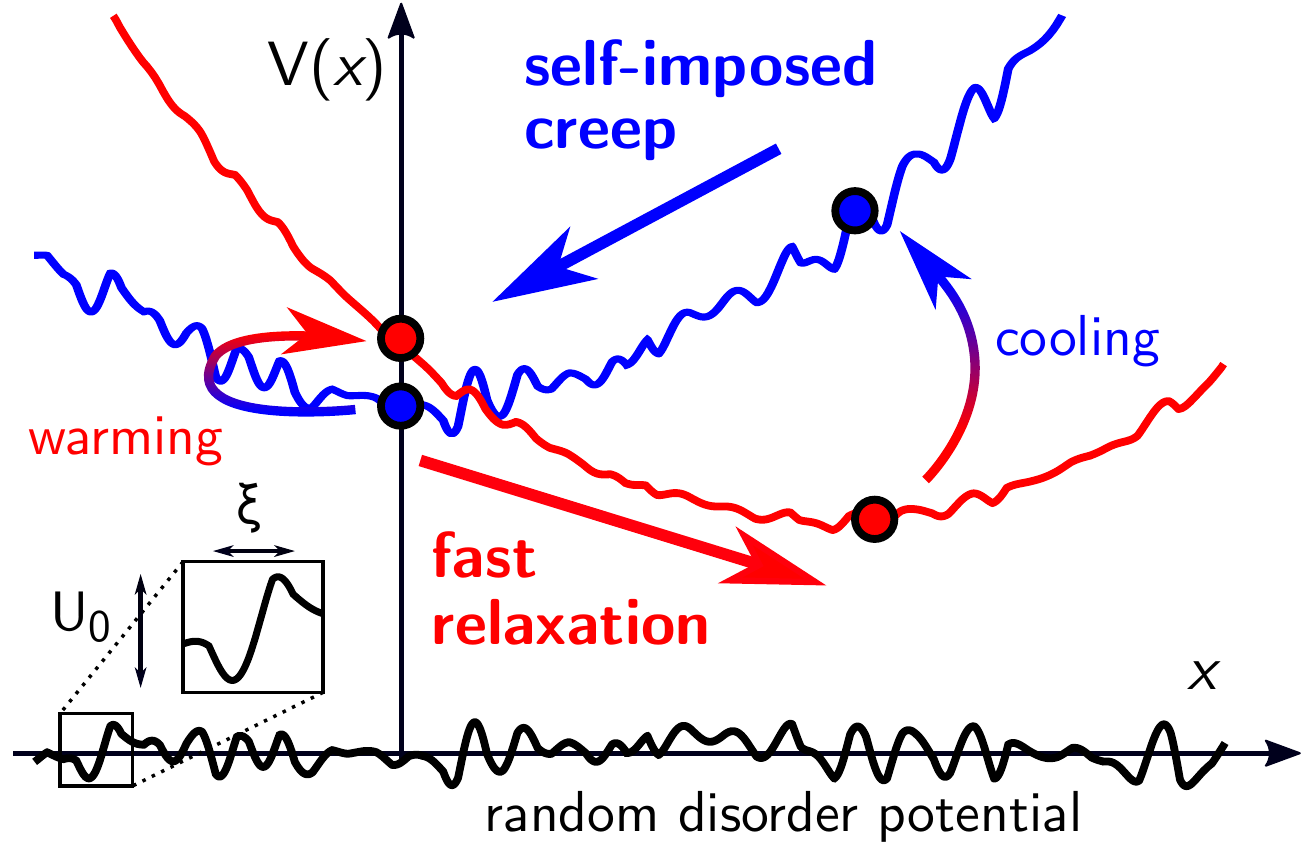}
\caption{Schematic response of a particle confined in a potential subject to random disorder. The disorder potential is characterized by local maxima of size $U_{0}$ and over a length scale $\xi$ [the superconducting coherence length in the case of vortices in superconductors]. Upon warming, the particle rapidly moves to a newly defined minimum as the disorder is thermally smeared out (indicated by effectively smaller wells, red). When cooling, the potential changes and the particle increases its energy relative to the new minimum in $V(x)$. As the motion is impeded by pinning barriers (blue), the particle creeps by thermal relaxation towards the new minimum.\vspace{-1em}
}
\label{fig:model}
\end{figure}

Superconducting vortices are lines of quantized magnetic flux $\Phi_{0} \!=\! hc/2e$ where the superconducting order parameter is depressed at a length scale of order of the superconducting coherence length $\xi$ and circular supercurrents are flowing at a length scale of order of the penetration depth $\lambda$. Vortices arrange in a lattice, which can be triangular, square or disordered due to the interactions of vortices with the crystalline environment\cite{Feigelman1989, Blatter1994, Brandt1995, Kogan1997, Guillamon2014}. Vortices tend to be pinned on material defects or inclusions consisting of places where superconductivity is depressed on length scales of order $\xi$. When varying the applied magnetic field, its strength or direction, vortices enter or exit the sample\cite{Feigelman1989, Blatter1994, Brandt1995}. This produces vortex motion, which is often counteracted by pinning. The action of the pinning landscape results in long-lived out-of-equilibrium vortex distributions that relax through thermal creep over a manifold of barriers\cite{Feigelman1989, Blatter1994, Brandt1995, Kierfeld2000, Dumont2002, Konczykowski2012, Klein2014, Herrera2017}. The phenomenon of creep has been observed in interacting systems of particles, such as colloids, polymers, solids consisting of mixtures or in lattices of entities formed through electronic interactions (domain walls or skyrmions)\cite{Bode2010,Landrum2016,Newland1982,DuttaGupta2015,Zhou2015}. Vortex lattices are often considered as a model system. Most efforts to understand vortex creep have focused on trying to immobilize vortices and thereby increase the critical current $j_{c}$ for applications\cite{Foltyn2007}. But the interaction between vortices and the underlying superconducting material is very rich and can produce counterintuitive phenomena. Here we find that the equilibrium state towards which the system creeps can be modified with the temperature, leading to a phenomenon that we term \emph{self-imposed creep}. We study vortex creep in the the superconductor $2\mathrm{H}$-$\mathrm{Nb}\mathrm{Se}_{2}$ and we observe that, as a consequence of self-imposed creep, the vortex lattice starts creeping when cooling, see Fig.\ \ref{fig:model}.

The layered system $2\mathrm{H}$-$\mathrm{Nb}\mathrm{Se}_{2}$ shows weak pinning and small creep rates. Vortex properties depend crucially on the magnetic field direction with respect to the layers. The axis normal to the layers provides a distinguished direction and one speaks about a uniaxial superconductor. The current distribution for a magnetic field tilted towards the layers yields a misalignment between the field and the flux-lines by angle that depends on temperature through the first critical field $\propto\! H_{c1}(T)$\cite{Clem1984, Giovannella1987, Liu1988, Blatter1994, Hasanain1999, Pal2006, Kogan2017}. The temperature induced variations in this misalignment angle modifies the equilibrium position for vortex creep.

\begin{figure}[tb]
\centering
\includegraphics[width=0.45 \textwidth]{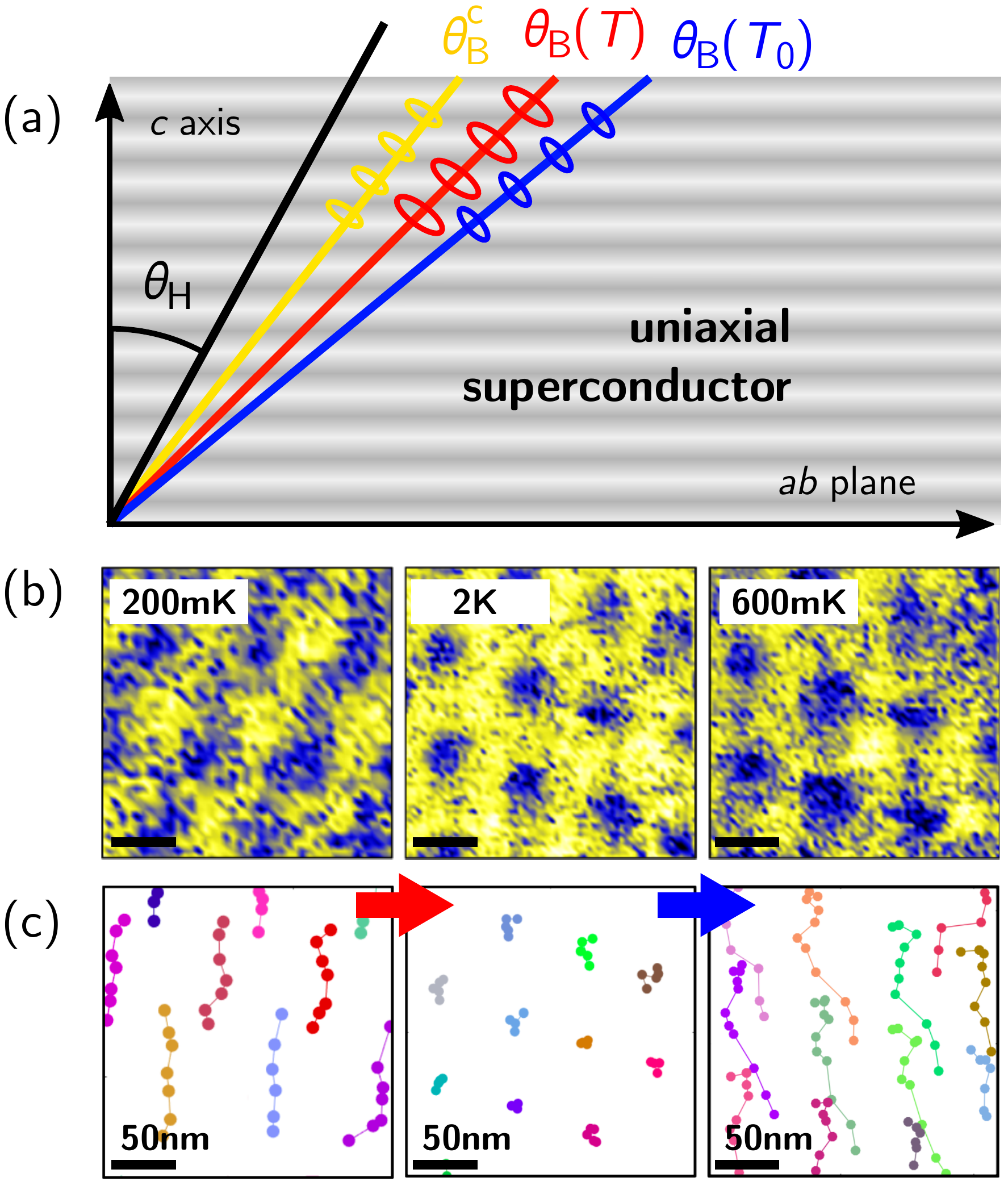}

\caption{%
(a) Schematic vortex alignment for anisotropic superconductor in tilted magnetic field ($\theta_{\sss H}$, black). The $T$-dependent current patterns---indicated as rings---define different equilibrium orientations $\theta_{\sss B}$ at low ($T_{0}$, blue) and higher ($T \!>\! T_0$, red) temperature. When tilting the magnetic field to $\theta_{\sss H}$, vortices creep towards the equilibrium angle $\theta_{\sss B}^{c} \!<\! \theta_{\sss B}(T_{0})$, yellow.
(b) Time averaged image of a series of STM images taken at three temperatures. At low $T$, out-of-equilibrium vortices move between subsequent frames. The motion stops upon warming to $2~\mathrm{K}$. At this temperature, vortices are fixed in a lattice. When cooling, motion reappears again and the average image shows that vortices do not stay at the same position in subsequent steps.
(c) Vortex positions extracted from each image (points). Each vortex is identified by a color and the lines join the position of each vortex in two consecutive images. We extract the relevant velocities in the Fig.\,\ref{fig:exp-velocity-jitter}.\vspace{-1em}
}
\label{fig:exp-displacement}
\end{figure}

To study vortex creep we use a dilution refrigerator Scanning Tunneling Microscope (STM) in a three axis vector magnet \cite{Galvis2015}. We use a gold tip, sharpened and cleaned in-situ \cite{Rodrigo2004b} and study a cleaved $2\mathrm{H}$-$\mathrm{Nb}\mathrm{Se}_{2}$ sample ($\sim 1 \!\times\! 1 \!\times\! 0.2~\mathrm{mm}^{3}$) grown with iodine vapor deposition. For the field strength $H \!=\! 0.85~\mathrm{T}$, the magnetization is practically reversible so that no vortex motion is measured within days. After field cooling to the base temperature ($T_{0} \!=\! 150~\mathrm{mK}$), we rotate the magnetic field from perpendicular to the layers ($c$ axis) towards the direction of the layers. We chose a tilt angle of 70$^{\circ}$, sufficient to install an out-of-equilibrium vortex state and observe creep while keeping tilt induced distortions in the symmetry of the vortex lattice or in vortex cores small. Distorted lattices in tilted fields in $2\mathrm{H}$-$\mathrm{Nb}\mathrm{Se}_{2}$ have been discussed in detail in
Refs.\ [\onlinecite{Hess1992, Hess1994, Gammel1994, Campbell1988, Fridman2011, Fridman2013, Galvis2018,Kogan1995}] and are related to the uniaxial anisotropy, $\varepsilon \equiv H_{c2,ab} / H_{c2,c} \approx 1/3$. The direction of the induction $\vec{B}$ differs from the direction of $\vec{H}$ by an angle $\theta_{\sss B} - \theta_{\sss H}$, with $\theta_{\sss H}$ and $\theta_{\sss B}$ being the angle of $\vec{H}$ and $\vec{B}$ relative to the $c$ axis. Minimizing the free energy with respect to $\theta_{\sss B}$ for fixed $H$ and $\theta_{\sss H}$ yields
\begin{align}\label{eq:angle}
   \sin(\theta_{\sss B} - \theta_{\sss H}) = \frac{H_{c1}}{H} \frac{(1-\varepsilon^{2})\sin\theta_{\sss B}\cos\theta_{\sss B}}{(\varepsilon^{2}\sin^{2}\theta_{\sss B} + \cos^{2}\theta_{\sss B})^{1/2}}
\end{align}
up to a logarithmic correction of order unity\cite{Balatskii1986, Kogan1988, Bulaevskii1991, Blatter1994}. The temperature dependence of the equilibrium angle $\theta_{\sss B}$, which depends on the current distribution and on $\lambda(T)$, is encoded in $H_{c1}(T)$. For small changes of $H_{c1}(T) = H_{c1}(T_{0}) + \delta H_{c1}$, large fields $H \gg H_{c1}$ and a tilt angle $\theta_{\sss B}$ away from 0 or $\pi/2$ around $T = T_{0}$, the angle changes to $\theta_{\sss B} (T) = \theta_{\sss B} (T_{0}) - \delta \theta_{\sss B}$, with
\begin{align}\label{eq:delta-theta}
   \delta \theta_{\sss B}
      &\approx - [\theta_{\sss B}(T_{0}) - \theta_{\sss H}]\, \delta H_{c1} / H_{c1}(T_{0}).
\end{align}
In the regime where pinning is weak, i.e., where the Bean length\cite{Bean1962} $\ell_{\sss B} \!\approx\! c B/4\pi j_{c}$ is larger than the sample width $w$ and thickness $d \lesssim w$, vortices in this critical state \cite{Bean1962, Willa2015b} are straight and oriented along the angle $\theta_{\sss B}^{c} \!=\! \theta_{\sss B} \!-\! (w/2 \ell_{\sss B}) \sin \theta_{\sss B} \!<\! \theta_{\sss B}$, see Fig.\ \ref{fig:exp-displacement}(a). With $j_{c} \sim 10^{3}~\mathrm{A/cm^{2}}$ this critical angle deviates from the equilibrium angle $\theta_{\sss B}$ as $\theta_{\sss B} \!-\! \theta_{\sss B}^{c} \approx 0.5^{\circ}$. Using $H_{c1}(T \!=\! 0) \!\approx\! 200~\mathrm{G}$ for the lower critical field, we find $\theta_{\sss B} \!-\! \theta_{\sss H} \!\approx\! 0.8^{\circ}$, i.e., vortices are more inclined towards the $ab$-plane than the external field, Fig.\ \ref{fig:exp-displacement}(a).  Finally, the Ginzburg-Landau scaling $H_{c1}(T) \approx H_{c1}(0)(1-T/T_{c})$, provides a relative change in $H_{c1}$ between the experiment's low ($T_{0} \!=\! 150~\mathrm{mK}$) and high ($T \!=\! 2~\mathrm{K}$) temperatures of $\delta H_{c1}/H_{c1} \approx  - 0.29$. This gives $\delta \theta_{\sss B} \approx 0.3^{\circ}$, a  non-negligible misalignment.

To measure vortex motion we make consecutive STM images as a function of time [each one taken in 23 minutes, the average over eight consecutive images is shown in Fig.\ \ref{fig:exp-displacement}(b)]. We extract the vortex displacements from the images and show these in Fig.\ \ref{fig:exp-displacement}(c). We note that vortices always move along the direction of the component of the magnetic field within the layers. We also note that vortex motion shows a weak modulation at distances which correspond to multiples of the inter-vortex distance. This self-matching effect has been reported earlier in $2\mathrm{H}$-$\mathrm{Nb}\mathrm{Se}_{2}$ and in disordered thin films \cite{Troyanovski1999, Troyanovski2002, Guillamon2011, Galvis2017} and evidences that the lattice moves as a whole in the creep regime. Other than that, no decay of vortex creep velocity is observed within the measurement time ($\sim\! 5~\mathrm{h}$). The vortex displacement ($160~\mathrm{nm}$) during this time translates to an angular velocity $10^{-3}~\mathrm{deg/h}$ of vortices tilting to new equilibrium positions, see Fig.\ \ref{fig:exp-displacement}(a). The large time scale for thermal decay is consistent with the misalignment between magnetic field and vortex axis discussed above, $|\theta_{\sss B} - \theta_{\sss B}^{c}| / \dot\theta_{\sss B} \sim 100~\mathrm{h}$. 

\begin{figure}[tb]
\centering

\includegraphics[width=0.45 \textwidth]{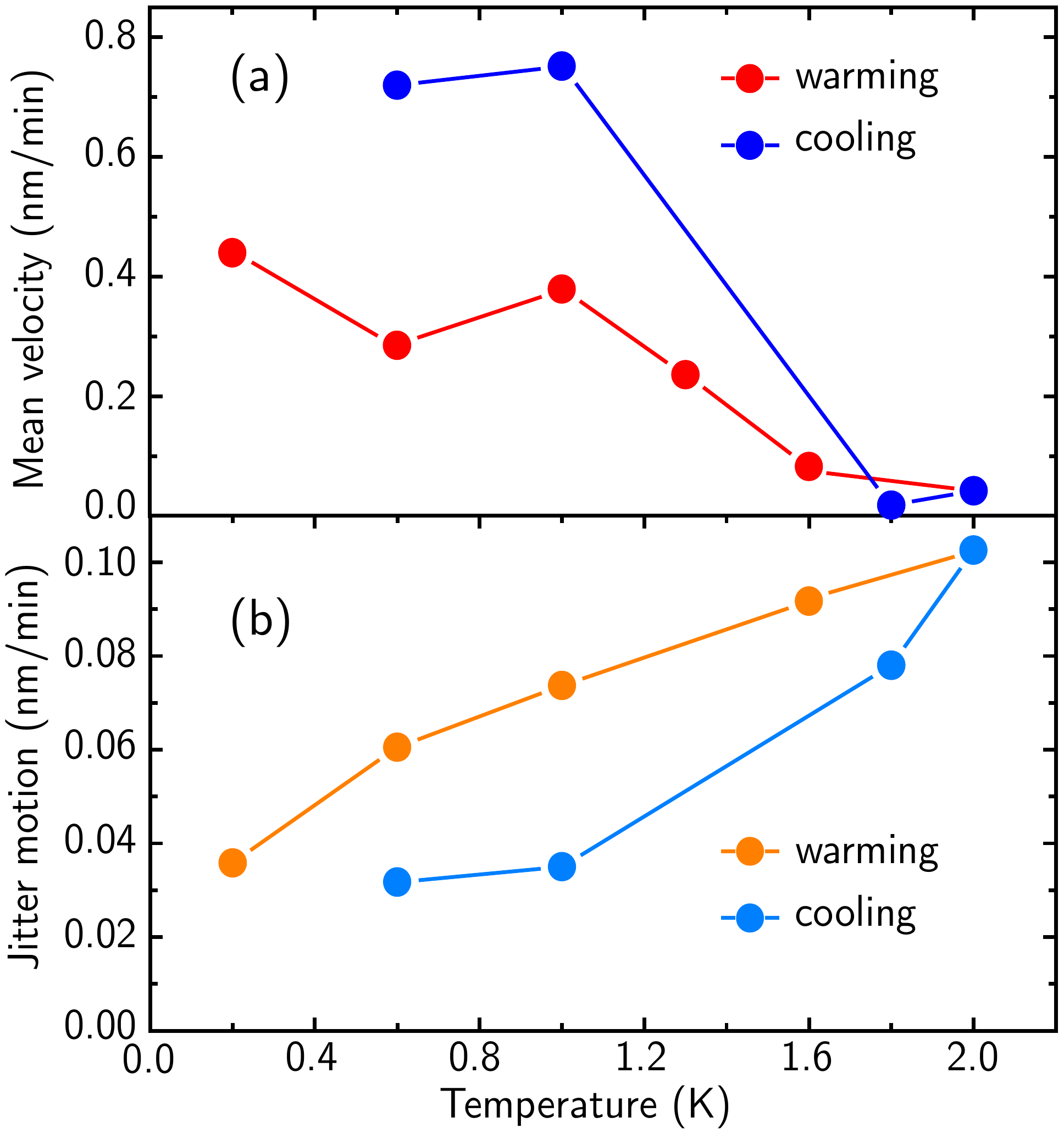}

\caption{
(a) Average velocity of vortices in $2\mathrm{H}$-$\mathrm{Nb}\mathrm{Se}_{2}$ at tilted field and fixed temperature, recorded on a warming (red) and cooling (blue).
(b) Standard deviation of the vortex jitter motion upon warming (orange) and cooling (light blue). The behavior agrees with the phenomenology of self-imposed creep.\vspace{-1em}}
\label{fig:exp-velocity-jitter}
\end{figure}

Imaging is repeated at different temperatures. It is important to stress here that no reinitialization occurs. Rather the system is kept at finite field strength and angle, and solely the temperature is changed. Hence it is temperature that determines the state preparation, whose relevance has been previously discussed in vortex physics\cite{Paltiel2000b, Willa2015a, Willa2018c}. In Figs.\ \ref{fig:exp-velocity-jitter}(a,b) we show the temperature dependence of the vortex velocity along the creep direction, the average creep velocity $v(T)$, and the vortex motion between consecutive images along random directions, the jitter motion $\Delta x (T)$, for a set of vortices, respectively. Each data point is obtained from a series of STM frames. To find $v(T)$ we determine the position $\vec{r}_{j}^{i}$ of vortex $j$ in frame $i$ and evaluate the average displacement for each vortex per frame, given by $\delta r_{j} = |\vec{r}_{j}^{n_{j}} \!-\! \vec{r}_{j}^{1}| / (n_{j} \!-\! 1)$, where $n_{j}$ denotes the number of frames where the $j$th vortex appears. Averaging over all $N_{v}$ vortices for a given temperature, we arrive at the average creep velocity $v(T)=\frac{1}{t_{\mathrm{f}} N_{v}} \sum\nolimits_{j=1}^{N_{v}} \delta r_{j}$
with $t_{\mathrm{f}} = 23~\mathrm{min}$ the time for measuring one frame. To quantify the jitter motion, we evaluate the average jitter displacement $\delta s_{j} \!=\! \big( \frac{1}{n_{j}-1} \sum\nolimits_{i=2}^{n_{j}} |\delta \vec{r}_{j}^{i}|\big) \!-\! \delta r_{j}$, with $|\delta \vec{r}_j^{i}|$ the vortex displacement between two subsequent frames $i\!-\!1$ and $i$. The average over all frames at a fixed temperature now provides the jitter motion $\Delta x (T)= \frac{1}{N_{v}} \sum\nolimits_{j=1}^{N_{v}} \delta s_{j}$.

The average creep velocity decreases upon warming and vanishes above $2~\mathrm{K}$. Upon cooling, however, a finite velocity reappears. If the vortices were to reach a temperature-independent minimum upon warming, the jitter motion would decrease upon cooling without a reappearance of creep motion. The reversible directed vortex motion upon thermal cycling is therefore a clear signature of self-imposed creep. 

\begin{figure}[t]
\centering
\includegraphics[width=0.42 \textwidth]{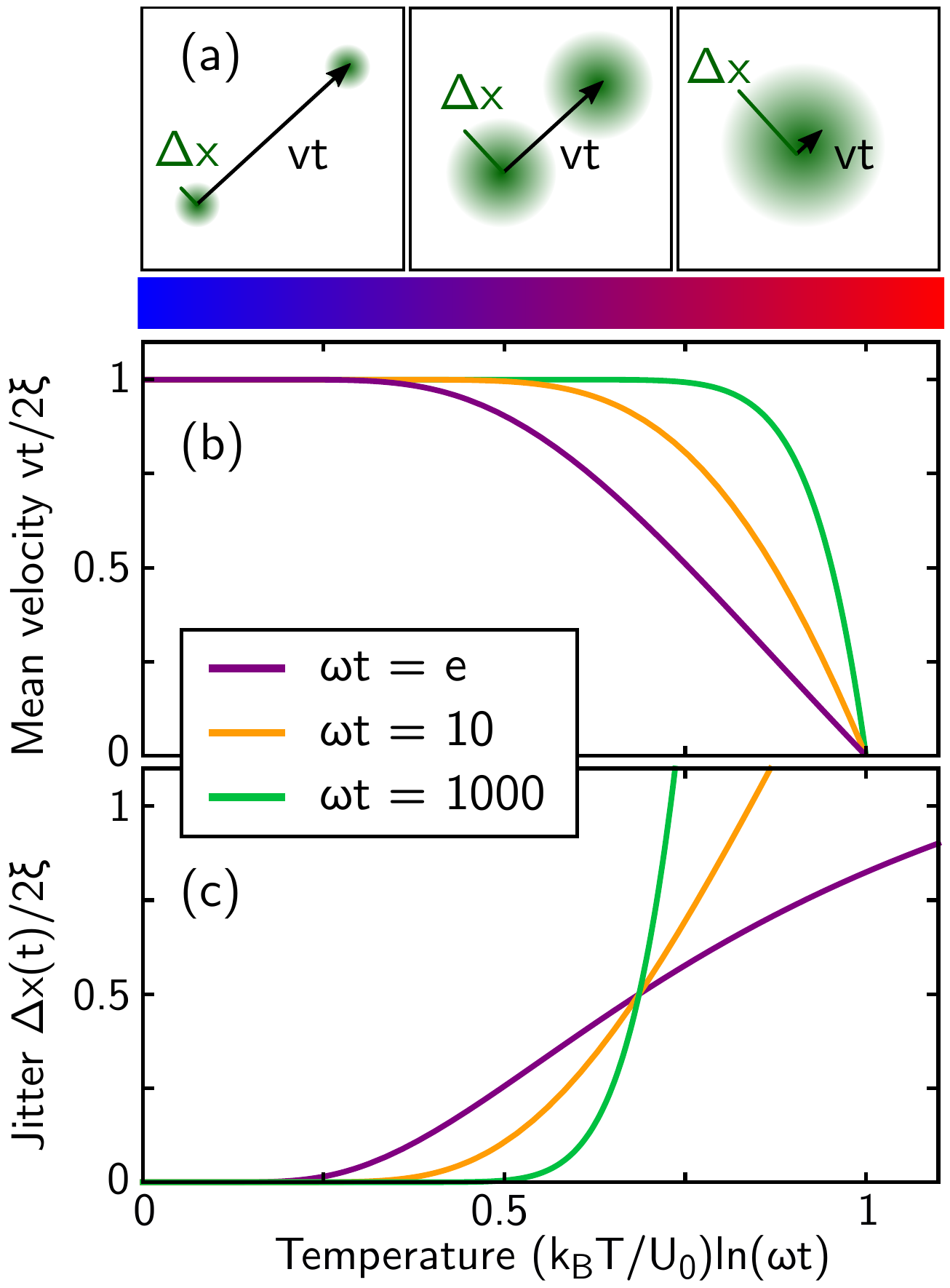}
\caption{
(a) Schematic of decreasing creep velocity (black arrow) and increasing jitter motion (green cloud) upon warming.
(b) Average creep velocity $v$ (in units of $2\xi/t$) as a function of temperature, see Eq.\ \eqref{eq:velocity-profile}, for different timescales $\omega t$. When temperature affects both the activation dynamics and the global potential minimum, see Fig.\ \ref{fig:model}, the velocity profile is traced reversibly upon warming and cooling.
(c) Standard deviation $\Delta x$ of the mean particle displacement, quantifying the isotropic thermal motion, or jitter, see Eq.\ \eqref{eq:bidirectional}.
\vspace{-1em}
}
\label{fig:velocity-profile}
\end{figure}

To capture the main observations we discuss a simple, yet quite generic, model for creep motion in a disordered landscape. First, let us note that motion is driven by thermal fluctuations, through an Arrhenius-type activation process across the pinning barriers. In such a case, the timescale $t \!\sim\! \tau \exp(U_{0}/k_{\sss B}T)$ for thermal activation is determined by (i) the temperature $T$, (ii) the energy barrier $U_{0}$, and (iii) a microscopic time scale $\tau \!=\! \omega^{-1}$ ($\omega$ is an attempt frequency to overcome the barrier).
Creep is observed when the time scale of the experiment is such that $U_{0} \!\sim\! k_{\sss B}T \ln (\omega t)$. To see the consequences of modifying the equilibrium, let us consider the problem of a particle confined in one-dimensional parabolic trap $V_0(x) = k x^{2}/2$ and subject to a driving force $V_d(x) \!=\! -F x$. In the context of a many-body system, the variable $x$ represents an observable parametrization, such as the angle between the magnetic field and the layers for the vortex lattice of a uniaxial superconductor.
The force produces a drive towards the equilibrium position $\bar{x} = F/k$. In addition,
let us consider
a disorder landscape $V_p(x)$ characterized by a typical depth $U_0$ and width $\xi$, with $k \xi^{2} / U_0 \!\ll\! 1$, see Fig.~\ref{fig:model}. We approximate the
bare potential between neighboring
minima $x_{\pm} \!=\! x \pm \xi$ by $V_{p}(x \!+\! \delta) \!\approx\! [1 - (\delta/\xi)^{2}] U_0$. The overall potential $V(x) \!=\! V_{0} + V_{d} + V_{p}$ features local minima in the range $\bar{x} - 2U_0/k\xi < x < \bar{x} + 2 U_0/k\xi$ and the position-dependent activation barrier $U_b(x)$ to move from $x_{-}$ to $x_{+}$ (we assume $x < \bar{x}$) is
\begin{align}\label{eq:barrier-of-x}
  U_{b}(x) = [(x-\bar{x})k \xi]^{2} / 4U_{0} + U_0 + (x-\bar{x})k \xi.
\end{align}
The thermally activated motion in the opposite direction, i.e. from $x_{+}$ to $x_{-}$ is penalized by an additional energy $-2 (x-\bar{x})k \xi > 0$. A particle initially far from the minimum $\bar{x}$ will glide down the potential until reaching $\bar{x} - 2U_0/k\xi$ from where it will be thermally activated across ever-growing barriers. After a time $t$, the particle has reached a position $x_{T}$ satisfying Arrhenius' condition
\begin{align}\label{eq:barrier-of-T}
  U_b(x_{T}) = k_{\sss B} T \ln(\omega t),
\end{align}
as smaller barriers have been overcome in exponentially shorter activation times.
Inserting Eq.\ \eqref{eq:barrier-of-x} into the condition \eqref{eq:barrier-of-T}, we obtain
\begin{align}
   x_{T} = \bar{x} - (2U_0/k \xi)\Big[1 - \sqrt{(k_{\sss B} T / U_{0}) \ln(\omega t)}\Big].
\end{align}
If $x_{T}$ is still far from $\bar{x}$ in the sense $\bar{x} - x_{T} \!\gg\! k_{\sss B} T \ln(\omega t)/k \xi$ [translating to $T \!\ll\! T_{b} \!\equiv\! (U_0/k_{\sss B})/\ln(\omega t)$], the particle moves with an average velocity $v \approx 2\xi/t$. If however the particle has relaxed in the vicinity of the global minimum, the thermal activation becomes almost equally probable in both directions. Accounting for this bidirectional, yet asymmetric, motion we find a net average creep velocity
\begin{align}\label{eq:velocity-profile}
\!\!
  v &\approx \frac{2\xi}{t} \Big\{1 - \exp\Big[-\!\frac{4 U_0 }{ k_{\sss B} T}\Big(1 - \sqrt{k_{\sss B} T \ln(\omega t) / U_{0}}\Big)\Big]\Big\}.\!
\end{align}
The temperature dependence of $v$ is shown in Fig.\ \ref{fig:velocity-profile}(a), for different values $\omega t$.
It is interesting to note that although the thermal activation energy increases upon warming, the creep velocity decreases.
The validity of the above result is limited to temperatures $T < T_{b}$. For larger temperatures, the disorder landscape becomes irrelevant, as the particle relaxes within $\sim\omega^{-1}$.

The directed relaxation via creep, Eq.\ \eqref{eq:velocity-profile}, is blurred by an isotropic contribution, or jitter motion with zero mean and amplitude (standard deviation) given by
\begin{align}\label{eq:bidirectional}
\!\!
   \Delta x \sim \xi \sqrt{\omega t}\exp\Big[-\!\frac{2 U_0 }{ k_{\sss B} T}\Big(1 - \sqrt{k_{\sss B} T \ln(\omega t) / U_{0}}\Big)\Big].\!
\end{align}
This result is obtained by describing the (forward-backward) activation across the barriers as a stochastic process (random-walk motion), where the variance $(\Delta x)^{2} = \xi^{2} \omega_{\mathrm{rw}}t$ of the displacement grows linearly in time and is determined by the random-walk attempt frequency $\omega_{\mathrm{rw}} = \omega \exp[2(x_{T}-\bar{x})k \xi/k_{\sss B}T]$. The exponent only depends on the difference between the barrier heights for forward and backward motion. This jitter motion persists beyond the disappearance of creep motion and reaches a similar magnitude when $k_{\sss B} T \ln(\omega t) \sim U_{0}$, as shown in Fig.\ \ref{fig:velocity-profile}(b).

Now consider what occurs if the temperature is modified. We have to distinguish two scenarios: one where the global minimum $\bar{x}$ is constant, and one where $\bar{x}(T)$ depends on $T$ through a temperature-dependent force $F(T)$.
In the first case, the average creep velocity is given by the local disorder landscape seen by the particle and hence follows Eq.\ \eqref{eq:velocity-profile}. For $T \!>\! T_{b}$, the system is fully relaxed and the particle reaches $\bar{x}$. Meanwhile the magnitude of the jitter motion continuously increases, see Fig.\ \ref{fig:velocity-profile}(b). Subsequent cooling lowers the thermal energy and the particle's motion freezes in place at $\bar{x}$. This is the conventional response expected for creep. In the second case, when $\bar{x}$ depends on $T$ through a temperature-dependent force $F(T)$, the situation is drastically different. $\Delta T$ imposes a shift $\bar{x}(T+\Delta T) - \bar{x}(T) \gg \xi$ of the global equilibrium position. Upon warming, the velocity and jitter motion are similar to the previous case. However, the values of $x(t)$ are different and, upon cooling, triggers particle motion tracing back (in magnitude) the warming curve, as schematically shown in Fig.\ \ref{fig:model}.

We compare our observations in $2\mathrm{H}$-$\mathrm{Nb}\mathrm{Se}_{2}$ with the above model. The typical barrier to overcome during pinning by thermal fluctuations is given by Arrhenius' law $U_{0} \!=\! k_{\sss B}T \ln(\omega t)$. To observe both the equilibrium phase at high temperature and reentrant creep at low temperature it is important that the temperature of the experiment is of order of $(U_{0}/k_{\sss B})/\ln(\omega t)$.
In contrast to the pinning energy of one defect site, here $U_{0}$ denotes the energy barrier for vortex creep \cite{Buchacek2018,Buchacek2019}. Weak collective pinning theory \cite{Larkin1979, Feigelman1989, Blatter1994} provides the estimate $U_{0} \!\sim\! k_{\sss B} T_{c} [(j_{c}/j_{\mathrm{dp}})(B/H_{c1})^{3}/Gi]^{1/2}$, with the depairing current $j_{\mathrm{dp}} \!=\! c \Phi_{0}/12\sqrt{3}\pi^{2} \lambda^{2}\xi$, the Ginzburg-Levanyuk \cite{Levanyuk1959, Ginzburg1960} number $Gi \!\sim\! [T_{c}/H_{c}(0)^{2}\xi^{3}]^{2}$, and the condensation energy $H_{c}(0)^{2}\xi^{3} \!=\! \Phi_{0}^{2}\xi/8\pi^{2}\lambda^{2}$. From Refs.\ [\onlinecite{Czapek1993, Banerjee1998, Menghini2002, Pasquini2008, Mohan2009, Maldonado2013}], we infer $j_{c}/j_{\mathrm{dp}} \!\sim\! 10^{-6}$ and $Gi \!\sim\! 10^{-4}$ and obtain $U_{0} \!\sim\! 10 k_{\sss B} T_{c}$. This estimate is compatible with $U_{0} \lesssim k_{\sss B} T_{c} \ln(\omega t)$, provided $\omega t \approx 2\times 10^4$; somewhat larger than the values considered above (Fig.\,\ref{fig:velocity-profile}). Given the simplicity of the one-dimensional model, the agreement is still remarkable. All important features predicted by the model---the disappearance and reappearance of the directed motion, together with the temperature-evolution of the jitter motion [behaviors of $v(T)$ and $\Delta x (T)$, shown in Figs.\ \ref{fig:exp-velocity-jitter}(a,b) and in Figs.\ \ref{fig:velocity-profile}(a,b)]---are found in the experiment.

Given that the experimental time scale spans several minutes, our observation $\omega t \!\sim\! 10^{4}$ suggest a value for $\omega$ of order of one Hz. While a route for accurate determination of this attempt frequency is still lacking, the estimate $\omega \!=\! \alpha_L/\eta$ for a single vortex depends on the vortex viscosity $\eta$ and on the Labusch parameter \cite{Labusch1969} $\alpha_L$ (which in turn relates to the averaged potential curvature\cite{Willa2016}). While values in the range $10^{6}$-$10^{10}~\mathrm{Hz}$ have been reported \cite{Brandt1989b}, the analysis assumes vibrations with large $k$-vectors. In our case, vortices are not isolated, but rather interact non-locally with many vortices \cite{Brandt1991, Blatter1994, Brandt1995, Yeshurun1996, Brandt2005}. Low $k$-vectors, or wave-lengths comparable to the sample size, leads to highly dispersive elastic moduli which modify the attempt frequency by orders of magnitude \cite{Yeshurun1996, Brandt2005, Feigelman1990b}. Similar to our observation, previous measurements of slow vortex dynamics have reported \cite{Raes2014, deSouza2016} very low frequency values for thermal motion and creep. Creep rates observed in layered cuprate superconductors involve extremely large time scales, indicating the relevance of collective creep \cite{Feigelman1990b, Blatter1994, Brandt1995, Yeshurun1996}. Thus, even if the attempt rate of individual vortices is large, the dynamics as a lattice involves rates that are many orders of magnitude smaller. The temperature is far from melting, thus favoring collective rather than a single-vortex dynamics \cite{Auslaender2009, Embon2015}. It is this near-equilibrium configuration with ultra-small collective dynamics that allows for the observed cooling imposed creep in our experiments.

The creep discussed here is very slow and no decay in the vortex velocity is observed within our experimental time. However, the creep rate $S = d \ln(j) / d\ln(t)$ can assume a seizeable value compatible with the suggested lower bound\cite{Eley2017} $S > (T/T_{c}) Gi^{1/2}$. Actually, $2\mathrm{H}$-$\mathrm{Nb}\mathrm{Se}_{2}$ is among the materials with lowest creep rates, close to MgB$_2$\cite{Eley2017}. Creep between metastable vortex states that occur near the order-disorder transition of the vortex lattice in $2\mathrm{H}$-$\mathrm{Nb}\mathrm{Se}_{2}$ or related to domain formation of lattices with different orientations in $\mathrm{Mg}\mathrm{B}_2$ has been reported\cite{Rastovski2013, Marziali2015, Marziali2017}. Motion then appears when modifying the relative strength of competing interactions, and it might well occur that the equilibrium configuration at some particular locations is influenced by temperature. Collective motion is also found in stochastic behavior of particle arrangements \cite{Bode2010}. Depending on particle interactions, the dynamics transits from individual random motion to flocking. The time scale related to flocking motion shows a divergent behavior with increasing interaction. Other long term dynamical behavior should appear in thermal effects and might lead self-imposed creep whenever there are two or more parameters influencing the behavior of the system. Mixtures, such as alloys, concrete or rocks \cite{Zhang2017}, liquids re-solidifying under stress \cite{Landrum2016}, steel under stress \cite{Newland1982}, colloidal systems, magnetic domain walls or skyrmions \cite{Zhou2015} are cases of complex systems where self-imposed creep may be induced from temperature-dependent interactions.

The model of self-imposed creep explains the critical state dynamics in $2\mathrm{H}$-$\mathrm{Nb}\mathrm{Se}_{2}$ at tilted magnetic fields; in particular the commonly unexpected appearance of vortex motion when cooling. Likely, the balanced thermal activation dynamics and the temperature-dependent equilibrium could be matched in other uniaxial superconductors with weak pinning and in complex systems.\medskip

\section*{Acknowledgements}
R.W.\ acknowledges the funding support from Swiss National Science Foundation (SNSF) through the Early PostDoc Mobility Fellowship.
J.B., E.H., I.G., and H.S.\ acknowledge support by the Spanish Research Agency (FIS2017-84330-R, MDM-2014-0377) and by the Comunidad de Madrid through program Nanomagcost-CM S2018/NMT-4321.
I.G.\ acknowledges support by EU European Research Council PNICTEYES grant agreement 679080.
E.H.\ acknowledges support from Departamento Administrativo de Ciencia, Tecnolog\'ia e Innovaci\'on, COLCIENCIAS (Colombia) Programa Doctorados en el Exterior Convocatoria 568-2012.
E.H.\ and J.A.G.\ acknowledge support from the Cl\'uster de Investigaci\'on en Ciencias y Tecnolog\'ias Convergentes (NBIC) de la Universidad Central (Colombia).
J.B., J.A.G., E.H., I.G., and H.S.\ acknowledge SEGAINVEX-UAM.
We wish to express a special thank to Gianni Blatter and Vadim B.\ Geshkenbein for enlightening discussions and stimulating the interpretation of the experiments, and to Christian Sp\r{a}nsl\"{a}tt for carefully reading the manuscript.

%

\end{document}